\documentclass[12pt]{article}
\usepackage{graphicx}

\newsavebox{\sboxpubnumber}
\newsavebox{\sboxpubdate}
\newcommand{\pubdate}[1]{\begin{lrbox}{\sboxpubdate}{#1}\end{lrbox}}
\newcommand{\pubnumber}[1]{\begin{lrbox}{\sboxpubnumber}{\begin{tabular}{l} #1 \\
				 \usebox{\sboxpubdate}
				 \end{tabular}}
                           \end{lrbox}
                           \pubblock}
\newcommand{\Title}[1]{\begin{center} {\Large #1 } \end{center}}
\newcommand{\Author}[1]{\begin{center}{ \sc #1} \end{center}}
\newcommand{\Address}[1]{\begin{center}{ \it #1} \end{center}}
\newcommand{\andauth}{\begin{center}{and} \end{center}}
\newcommand{\pubblock}{\rightline{
			\usebox{\sboxpubnumber}}}
\newenvironment{Abstract}{\begin{quotation}  }{\end{quotation}}
\newenvironment{Presented}{\begin{quotation} \begin{center}
             PRESENTED AT\end{center}\bigskip
      \begin{center}\begin{large}}{\end{large}\end{center}
      \end{quotation}}
\newcommand{\Acknowledgements}{\bigskip  \bigskip \begin{center} \begin{large}
             \bf ACKNOWLEDGEMENTS \end{large}\end{center}}

\def\al{\alpha}  
 
\def\ga{\gamma}

\def\si{\sigma}

\def\bk{{\mathbf{k}}}

\def\bu{{\mathbf{u}}}

\def\bA{{\mathbf{A}}}
\def\bB{{\mathbf{B}}}
\def\bJ{{\mathbf{J}}}

\def\bna{\mbox{\boldmath $\na$}}
\def\bdot{\mbox{\boldmath $\cdot$}}

\newcommand{\ben}{\begin{equation}}
\newcommand{\een}{\end{equation}}
\newcommand{\bea}{\begin{eqnarray}}
\newcommand{\eea}{\end{eqnarray}}
\newcommand{\ba}{\begin{array}}
\newcommand{\ea}{\end{array}}
\newcommand{\bi}{\begin{itemize}}
\newcommand{\ei}{\end{itemize}}

\def\math{\mathsurround 0pt}
\def\oversim#1#2{\lower.5pt\vbox{\baselineskip0pt \lineskip-.5pt
        \ialign{$\math#1\hfil##\hfil$\crcr#2\crcr{\scriptstyle\sim}\crcr}}}
\def\lap{\mathrel{\mathpalette\oversim {\scriptstyle <}}}

\def\pa{\partial}

\def\na{\nabla}

\begin{document}

\begin{titlepage}
\pubdate{\today}                    
\pubnumber{SUSX-TH/02-004 \\ NORDITA-2002-2 AP} 

\vfill
\Title{MHD inverse cascade in the early Universe}
\vfill
\Author{Mark Hindmarsh\footnote{email: m.b.hindmarsh@sussex.ac.uk}}
\Address{Centre for Theoretical Physics, University of Sussex, 
Brighton BN1 9QJ, U.K.}
\vfill
\andauth
\vfill
\Author{M. Christensson\footnote{email: mattias@nordita.dk}, 
A. Brandenburg\footnote{brandenb@nordita.dk}}
\Address{Nordita, Blegdamsvej 17, DK-2100 Copenhagen \O, Denmark}
\vfill
\begin{Abstract}
 We have carried out numerical simulations 
 of freely decaying magnetohydrodynamic (MHD) turbulence in three 
 dimensions, which can be applied to the evolution of stochastic 
 magnetic fields in the early Universe.
 For helical magnetic fields an inverse cascade effect is observed in 
 which magnetic helicity and energy is transfered from smaller scales to larger scales,
 accompanied by power law growth in the characteristic 
 length scale of the magnetic field. 
 The magnetic field quickly reaches a scaling regime 
 with self-similar evolution,
 and power law behaviour at high wavenumbers. We also find power law decay 
 in the  magnetic and kinematic energies. 
\end{Abstract}
\vfill
\begin{Presented}
    COSMO-01 \\
    Rovaniemi, Finland, \\
    August 29 -- September 4, 2001
\end{Presented}
\vfill
\end{titlepage}
\def\thefootnote{\fnsymbol{footnote}}
\setcounter{footnote}{0}


\section{Introduction}

Magnetic fields are ubiquitous in the Universe, being observed 
in nature on scales  
from planetary size to galaxy cluster size \cite{Zel88,Kro94}. 
In galaxies and galaxy clusters, the typical 
strength is of order a few $\mu$Gauss, which is thought to be produced by 
dynamo action on a seed field. 
In galaxies the dynamo timescale is roughly a rotation 
period, $10^8$ Yr, and a simple calculation \cite{Zel88}
based on the age of a typical 
galaxy shows that the seed field must have been about 
\(10^{-20}\) Gauss, or perhaps less in the 
currently favoured models with a cosmological term \cite{Davis:1999bt}.

There is no shortage of ideas for generating this seed field.  The more 
conventional astrophysical explanations are based on a Biermann battery 
operating at the era of reionisation (see e.g.\ \cite{GneFerZwe00} and the 
references therein).  There are more speculative ideas based on various 
models of inflation \cite{InfMag}, phase transitions \cite{PhaTraMag} and 
primordial black holes \cite{BlaHolMag}.

All these mechanisms have the common feature of producing stochastic, 
homogeneous and isotropic magnetic and velocity fields which can be 
characterised by their power spectra and characteristic initial scales. Our 
interest here is to try and make model-independent statements about the 
evolution of the magnetic fields once they are generated.  This 
article, which is based on Ref.~\cite{ChrHinBra01}, studies the evolution of a 
stochastic magnetic field generated at a phase transition, such as the 
confinement transition in QCD at $t \simeq 1$ sec, or the electroweak 
symmetry-breaking transition at $t \simeq 10^{-11}$ sec.  It therefore 
falls into the category of decaying 3D MHD turbulence, which has been 
studied before in the MHD community
\cite{Hos+95,PolPouSul95,GalPolPou97,Mac+98,BisMul99,MulBis00}. 
Most directly comparable to our work, 
Biskamp and M\"uller \cite{BisMul99} studied the energy decay in incompressible
3D magnetohydrodynamic turbulence in numerical simulations at relatively high 
Reynolds number, and in a companion letter \cite{MulBis00} studied the 
scaling properties of the energy power spectrum.

We focus here on the transfer of magnetic energy from small to 
large scales, as necessitated by the conservation of magnetic helicity. 
This is important for a primordial magnetic field to reach a 
large enough scale with sufficient amplitude to be relevant for seeding the 
galactic dynamo \cite{HinEve98}.

We perform 3D simulations both with and without magnetic helicity, 
starting from statistically homogeneous and isotropic random initial 
conditions, with power spectra suggested by cosmological applications. 
We find a strong inverse cascade in the helical case, with the coherence 
scale of the field growing as $t^{0.5}$, and equivocal 
evidence for a weak cascade when only helicity fluctuations are present. 
In the helical case we also find a
self-similar power spectrum with an approximately $k^{-2.5}$ behaviour at 
high $k$.
We find decay laws for the magnetic and kinetic energies of 
$t^{-0.7}$ and $t^{-1.1}$, respectively, in the helical case, and $t^{-1.1}$ for both in 
the non-helical case.

\section{Evolution of magnetic fields in the early Universe}

A convenient benchmark is to assume that 
the field is created on the horizon scale with a power spectrum $k^n$, 
taking all the energy in the Universe, and that it is subsequently 
completely frozen into the plasma \cite{HinEve98}.  If one takes 
the epoch of creation to be either the QCD transition or the electroweak 
transition, one finds that the RMS fluctuations on the scale of a protogalaxy 
are roughly
\[
B(t_0,L=0.5 \mathrm{Mpc})
< \left\{ \ba{ll} 10^{-14-3n}\; {\textrm G} &
\mathrm{QCD} \\[-2pt]
10^{-20-4.5n}\; {\textrm G} & \mathrm{EW} \ea \right.
\]
Thus we see that for causal fields ($n \ge 2$) there needs to be some 
amplification on large scales for there to be any interesting seed field for 
the galactic dynamo.

In fact, the field is not strictly frozen into the plasma, and in order to calculate 
observable effects, we must determine how the field scale length
\(\xi\) and magnetic energy \(E_M\) evolves.  There are various scaling 
arguments which have been put forward: for example, for ideal MHD (infinite 
conductivity) and with no helicity Olesen \cite{Ole97} (and later 
Son \cite{Son:1999my}, Field and Carroll \cite{Field:2000hi} and Shiromizu 
\cite{Shiromizu:1998bc}) argued
\ben
\xi(t) \sim t^{2/(n+5)},
\een
where $t$ is conformal time.  The effect of having a conserved helicity 
modifies this scaling law to \cite{Bis93, Field:2000hi,Son:1999my}
\ben
\xi(t) \sim t^{2/3},\qquad E_M \sim t^{-2/3}.
\een
Early numerical experiments with a shell model of the full MHD equations 
\cite{BraEnqOle96} suggested \(\xi \sim t^{0.25}\), and focussed attention 
on the possibility of an inverse cascade, in which power is transferred 
locally in $k$-space from small to large scales.

\section{MHD equations}

The MHD equations in an expanding Universe are most conveniently expressed 
in terms of conformally rescaled fields $\bB$, $\bu$ and dissipation parameters 
$\nu$, $\eta$, and in the gauge $A^0 = \eta \bna\bdot\bu$ 
\cite{Subramanian:1998gi,ChrHinBra01}.

The matter and radiation in the early Universe is modelled as an 
isothermal compressible gas with a 
 magnetic field, which is governed by the momentum equation, the continuity
 equation, and the induction equation, written here in the form
 \bea
    \frac{\pa\bu}{\pa t} &=& 
    -\bu\cdot\bna\bu - 
    c_{s}^{2}\bna \ln\rho +
    \frac{\bJ\times\bB}{\rho}
    +\frac{\mu}{\rho}\left(\na^{2}\bu + 
    \frac{1}{3}\bna\bna\cdot\bu\right),\\
    \frac{\pa \ln\rho}{\pa t} &=& 
    -\bu\cdot\bna \ln\rho - 
    \bna\cdot\bu,\\
    \frac{\pa\bA}{\pa t} &=& 
    \bu\times\bB + 
    \eta\na^{2}\bA,
\eea
 where $\bB=\nabla\times \bA$ is the magnetic field in terms of the magnetic
 vector potential $\bA$, $\bu$ is the velocity, $\bJ$ is the current density, 
 $\rho$ is the density, $\mu$ is the dynamical viscosity, and $\eta$ is the 
 magnetic diffusivity.
 
In the ideal limit $\mu=\eta=0$, there is a conserved quantity in addition to 
the energy, which is the magnetic helicity $H_{\rm M}$, given by
\ben
 H_{\rm M} = \int\bA\cdot\bB\, d^{3}x.
\een
Helicity is known to be important in dynamo theory 
\cite{PouFriLeo76,MenFriPou81},
where turbulence is driven. We shall also be able to confirm its 
importance in decaying turbulence.
  
An important dynamical quantity is 
 the magnetic Reynolds number $Re_{\rm M} = Lv/\eta$,
 where $L$ and $v$ are the typical length scale and velocity of the
 system under consideration, because it  measures the importance of the 
 non-linear  term in the equation for the magnetic field.
 In the early Universe, $Re_{\rm M}$ can be very large. 
 This is often taken to mean that the  magnetic field is frozen into 
 the plasma, and the scale length of the field increases only with the 
 expansion of the Universe. This is in general untrue, because turbulence 
 can transfer energy to different length scales \cite{BraEnqOle96}. 
 Not only can there be the usual direct cascade of energy from large to 
 small scales,  but also an inverse cascade, 
 increasing the overall comoving correlation length \cite{PouFriLeo76}.
 
\section{Plasma properties in the early Universe}

We are considering the time before recombination, 
when the plasma in the early Universe has a 
relativistic equation of state $p=\rho/3$, and therefore a sound speed 
$c_s = 1/\sqrt{3}$.  The number species contributing to the pressure 
gradually decreases with the temperature $T$, until $T\simeq m_e$, the 
electron mass, when only photons and neutrinos remain relativistic.  At that 
point the number of charge carriers in the plasma reduces by a very large 
factor: for $T \gg m_e$, the electron number density $n_e$ is approximately 
equal to the photon number density $n_\gamma$, while for $T \ll m_e$ 
$n_e/n_\gamma = n_B/n_\gamma \simeq 10^{-10}$, where $n_B$ is the baryon 
number density. 

The transport properties of the plasma are determined from the mean free path 
$l_{\rm mfp}$
of the relevant particles involved in the transport of the quantity of 
interest, which typically is \(l_{\rm mfp} \sim 1/\al|\log\al|T \) 
\cite{TraCoe}, where 
$\al$ is the fine structure constant. From this we can infer 
that the conductivity is 
\ben 
   \si \sim 
        \left\{\ba{ll}T/\al|\log\al| & T \gg m_e \\
            (n_e/n_\ga)(m_e^2/T)/\al &  T \ll m_e \\
            \chi(n_B/n_\ga)(T/m_e)^2Te^{-2} & T < T_{\mathrm{dec}} 
         \ea\right.,
\een
where $T_{\mathrm{dec}}$ is the temperature at photon decoupling, roughly 1 eV.

The viscosity parameter $\nu$ is given by
\ben 
   \nu \equiv \eta/\rho \sim 
   \left\{ \ba{ll} 1/\alpha^2|\log\alpha|T  &  T \gg m_e\\
                (n_\gamma/n_e)(m_e^2/T^2)/\alpha T &  T \ll m_e
           \ea
   \right.
\een

We can estimate the importance of non-linear terms 
relative to diffusion from the Reynolds numbers of the plasma,  the 
hydrodynamical 
\( \mathrm{Re} = vL/\nu \) and magnetic \( \mathrm{Re}_M = vL\si\).
Upper bounds are obtained by assumed a fluid flowing at the speed of light 
on a scale somewhere between the mean free path and the particle horizon $ct$:
\ben 
(\al,1/\al^2\log^2\al) \lap (\mathrm{Re},\mathrm{Re}_M)
\lap 10^{19} (T/\mathrm{GeV})^{-1}
\een

\section{3D MHD simulations of decaying turbulence}

We solve these equations numerically with a code \cite{Bra01}
 using a variable third order Runge-Kutta
 timestep and sixth order explicit centred derivatives in space. 
 All our runs are performed on a $120^3$ grid, and 
 we use periodic boundary conditions,
 which means that the average plasma density
 $\langle \rho_{0} \rangle = \rho_{0}$ is conserved during runs.
 Here $\rho_{0}$ is the value of the initially uniform density, and the 
 brackets denote volume average. 
 
We use natural units $c=1$, and set the unit of length by setting 
measure $k_{1}=1$, where $k_{1}$
is the smallest wave number in the simulation box.  Hence the 
box has a size of  $L_{\rm BOX} = 2\pi$. 
The scale factor is fixed by setting 
$\rho_0=1$, and $\bB$ is measured
in units of $\sqrt{\mu_{0}\rho_{0}} c$, where $\mu_{0}$ is the magnetic
permeability. 
We define the mean kinematic viscosity $\nu \equiv \mu/\rho_{0}$.
The sound speed $c_{s}= 1/\sqrt{3}$, as appropriate for a relativistic fluid.

The equations are not quite those for a 
relativistic gas in the early universe \cite{BraEnqOle96}. However, we have 
checked that our results change little when using the true relativistic 
equations in the low velocity limit.
 
The initial conditions appropriate for the early Universe are to take both the 
velocity and magnetic fields to be homogeneous and isotropic Gaussian random 
fields drawn from a power-law distribution with a high wavenumber cut-off, 
determined by the physical mechanism generating the power spectrum. 
The power spectra, defined by  
 $P_{\rm M}(k)\equiv \langle\bB^{*}_\bk\cdot\bB_\bk\rangle$, and
 $P_{\rm V}(k)\equiv \langle\bu^{*}_\bk\cdot\bu_\bk\rangle$
is taken to have the form
\ben
P_{\rm M}(k)=A_M k^{n} \exp ( - (k/k_c)^4),\quad
P_{\rm V}(k)=A_V k^{m} \exp ( - (k/k_c)^4).
\een
note that in the plots it is the 
shell-integrated energy spectra, 
$E_{\rm M,V}=4\pi k^2\times{1\over2}P_{\rm M,V}(k)$,
which are shown.

Note also that causality demands that $n\ge 2$ and 
$m\ge 0$ \cite{DurKahYet98}.
In the simulations presented we took the power laws to be the lowest 
consistent with causality.  We also chose $k_c = 30$, 
unless specified otherwise. The initial magnetic energy was taken 
equal to the kinetic energy, and  had the value $5\times10^{-3}$ in all runs.
  
We are also able to control the helicity in the initial conditions, 
ranging from identically zero to maximal, where maximal helicity is 
defined as saturating the inequality
\ben
   | H_{\rm M}(k)|\leq 2k^{-1} E_{\rm M}(k).
\een
It should be noted that if one chooses the initial field with no constraint, 
there are helicity fluctuations present, around a mean of zero.  It 
is possible to completely eliminate the fluctuations but in practice they 
are quickly regenerated by the evolution of the field.

\section{Results}
In all runs the mean kinematic viscosity $\nu$ and the resistivity $\eta$
were chosen to be equal 
with values between $\nu=\eta=5\times10^{-4}-5\times10^{-5}$.

\subsection{Magnetic energy spectrum}
\label{s:specev}

\begin{figure}[h]              
  \centering
  \scalebox{0.38}{\includegraphics{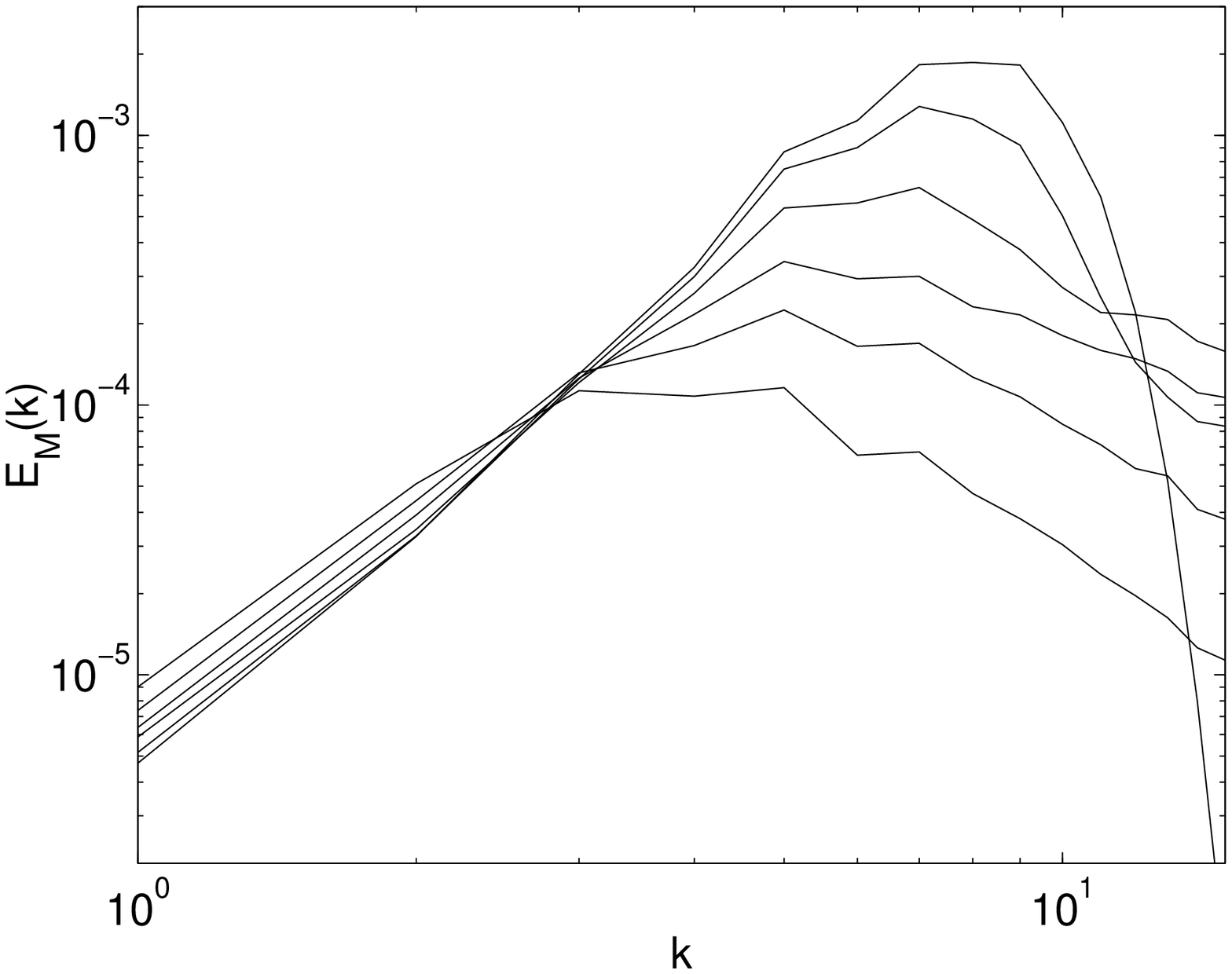}}
  \scalebox{0.38}{\includegraphics{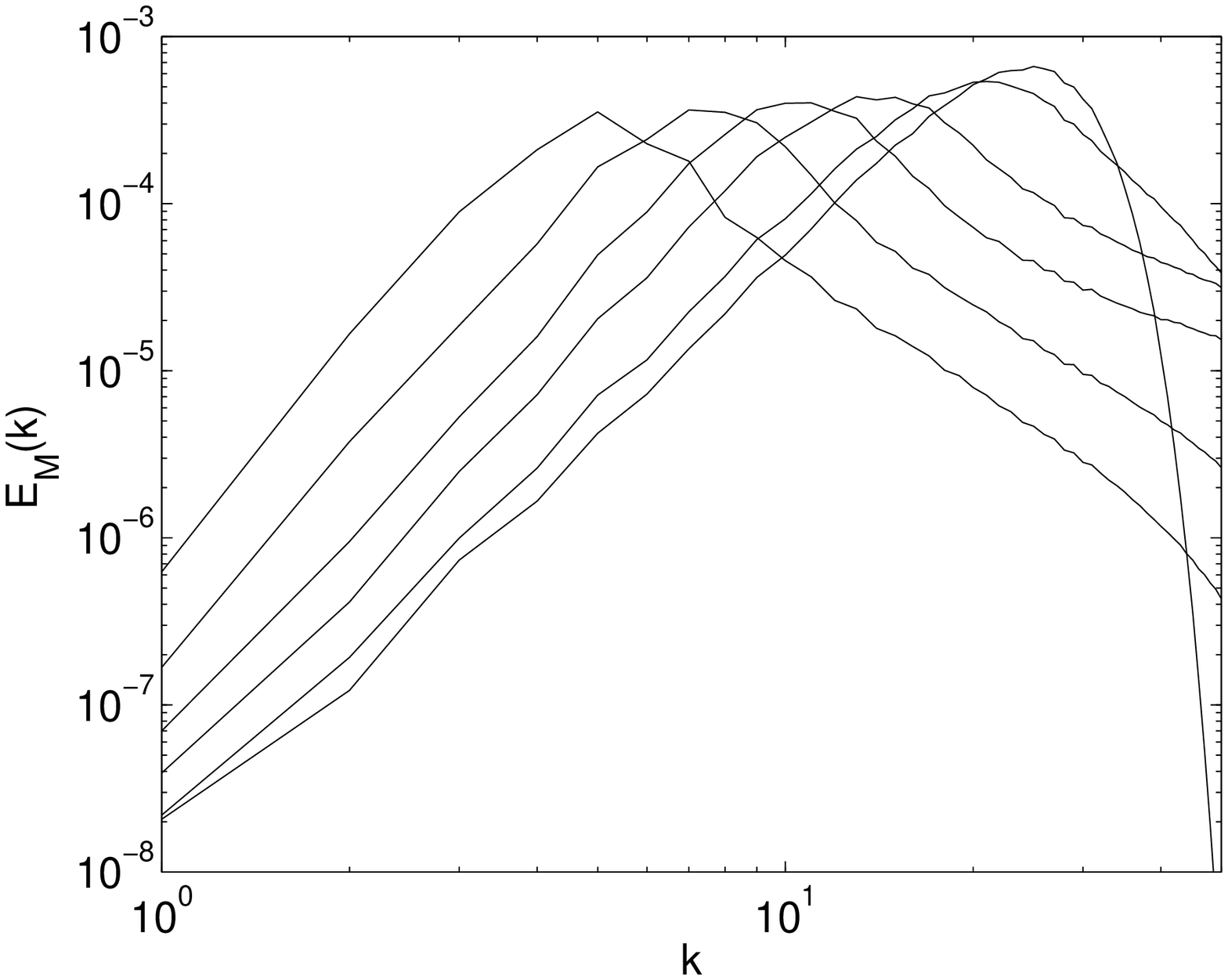}}  
  \caption{Magnetic energy spectrum $E_{\rm M}(k)$ for a run with unconstrained 
  (left) and maximal (right) magnetic 
  helicity. $\nu=\eta=5\times 10^{-5}$. The times shown are 
  $0, 1.0, 4.6, 10.0, 21.5$ and $46.3$. At low wavenumbers $k$ the energy 
  spectrum
  $E_{\rm M}(k)$ increases with time.}
  \label{spec.mag}
\end{figure}

In Fig.~\ref{spec.mag} we show magnetic energy spectra $E_{\rm M}(k)$ for 
runs with unconstrained and maximal initial magnetic helicity.
%
%
%
%
%
Turning first to maximal initial helicity, one sees that 
after a short initial direct cascade, 
there is energy transfer towards lower $k$, and the peak of the power 
spectrum moves to the left, signifying a growth in the coherence scale of
the field.  This has been called an inverse cascade, although it should be 
noted that the term ``cascade'' carries the implication of a local interaction 
in $k$-space, which is not necessarily true 
\cite{AckJens}.  In fact, in simulations where 
the initial conditions contain power in only a single wavenumber, the $k^4$ 
spectrum is very quickly established at low $k$ \cite{ChrHinBra02}, 
indicating that the power transfer to large scales is non-local in $k$-space \cite{Bra01}.

For wavenumbers above the peak, one sees a decrease in power with quite a 
different form than is present in the initial conditions, which can be fitted 
approximately by $E_{\rm M}(k) \sim k^{-2.5}$.  

When only helicity fluctuations are present in the initial conditions, the 
peak in the power spectrum still moves to low $k$, but there is very little 
transfer of power.  It is perhaps surprising to see any transfer at all, as 
there are arguments which connect the evolution of the coherence length 
with helicity conservation \cite{Bis93,Field:2000hi}.  In an attempt to test this 
idea we also did a run with identical parameters but zero initial helicity, 
but saw essentially identical behaviour, with a small increase in power at
small scales.

In our runs with $E_{\rm K} = E_{\rm M}$ initially, the kinetic energy 
spectrum shows 
no evidence of an inverse cascade at any scale. 
However, when the initial velocity distribution
is zero the kinetic spectrum grows on all scales initially and in the low 
wave number region the energy continue to grow even after the high wave number 
modes start to decay.

\subsection{Coherence length evolution}
\begin{figure}[h]                          
  \centering
  \scalebox{0.475}{\includegraphics{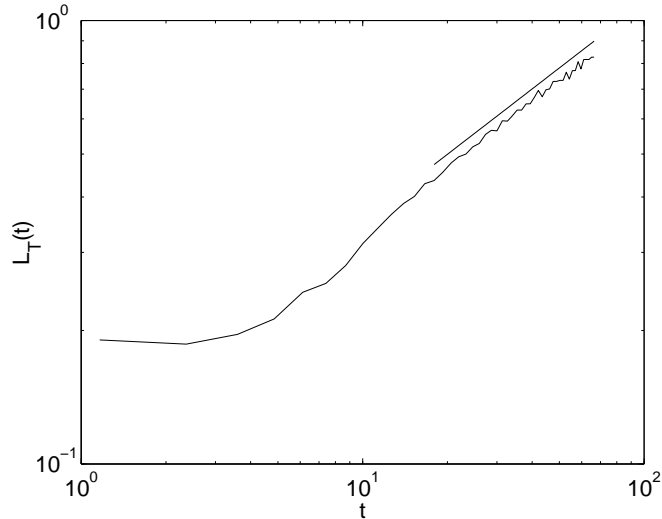}}
  \caption{Time evolution of the magnetic Taylor microscale 
    for with maximal initial magnetic helicity. $\nu=\eta=5\times 10^{-5}$.
    The straight line indicates the power law $\propto t^{0.5}$.}
  \label{taylor_hel}
\end{figure}

One length scale is the magnetic Taylor microscale
$L_{\rm T}$, defined as $B_{\rm rms}/J_{\rm rms}$. 
In Fig.~\ref{taylor_hel} we show $L_{\rm T}$ against time $t$
for a run with maximal initial helicity.
%
%
%
%
%
The asymptotic behaviour of the length scale is seen to grow 
approximately as $L_{\rm T}\sim t^{0.5}$. 


In runs with non-helical initial conditions the growth of the magnetic Taylor
microscale is slower: we find approximately 
$L_{\rm T}\sim t^{0.4}$.

\subsection{Energy decay laws}
\begin{figure}[h]
\centering              
\scalebox{0.475}{\includegraphics{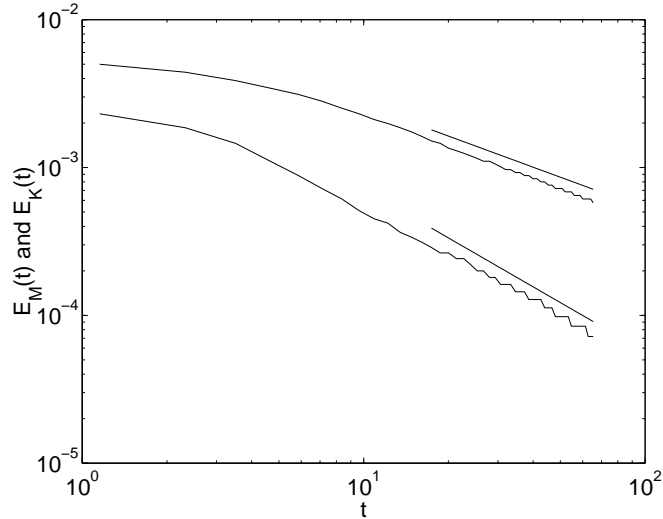}}
\caption{Time evolution of the magnetic energy $E_{\rm M}(t)$ and 
the kinetic energy $E_{\rm K}(t)$ in the case where there is 
initial magnetic helicity. $\nu=\eta=5\times 10^{-5}$.
The straight lines indicate the power laws $\propto t^{-0.7}$ and 
$\propto t^{-1.1}$ respectively.}
\label{brms_hel}
\end{figure}

Figure~\ref{brms_hel} shows the magnetic  
energy $E_{\rm M}(t)$ and the kinetic energy $E_{\rm K}(t)$ with 
maximal initial helicity.
%
%
%
%
%
It is seen that the asymptotic decay rate for $E_{\rm M}(t)$ is 
approximately $t^{-0.7}$. 
The kinetic energy also decays with a power law behaviour at late times: 
$E_{\rm K}(t) \sim t^{-1.1}$. 

In runs without initial helicity the decay rates of $E_{\rm M}(t)$ and
$E_{\rm K}(t)$ are approximately the same, close to $t^{-1.1}$.

\subsection{Magnetic Reynolds number}

\begin{figure}[h]
\centering
\scalebox{0.475}{\includegraphics{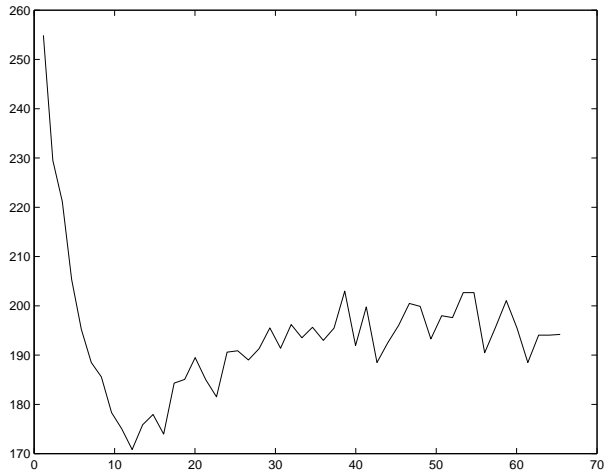}}
\caption{\label{f:mag.rey} Magnetic Reynolds number for the run of Fig.\ 
\ref{spec.mag}}
\end{figure}

The Reynolds numbers in our simulations are 
evaluated using the magnetic Taylor microscale.  
In most our simulations we typically obtain Reynolds numbers 
of the order of $100-200$. In Fig.\ \ref{f:mag.rey} we show the magnetic 
Reynolds number of the run with maximal helicity whose power spectrum is 
shown in Fig. \ref{spec.mag}.
%
%
%
%
%
Note that for the second half of the run the magnetic Reynolds number is 
approximately constant.

\subsection{Self-similarity in magnetic power spectrum}

We make the following ansatz for the energy spectrum
\ben
E_{\rm M}(k,t) = 
\xi(t)^{-q} g_{\rm M}(k\xi)
\label{self-similar},
\een
where $\xi$ is the characteristic length scale of the magnetic field,
taken to be the magnetic Taylor microscale defined above, and $q$ is a 
parameter whose value is some real number.
%
%
\begin{figure}[h]              
   \centering
   \scalebox{0.38}{\includegraphics{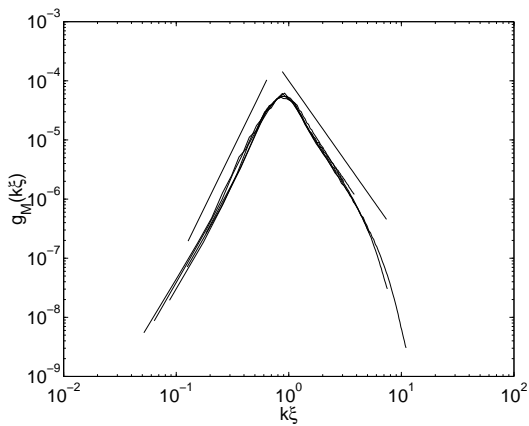}}      
   \caption{The magnetic scaling function $g_{\rm M}(k\xi)$ described in 
      the text, equation (\ref{self-similar}), against $k\xi$, with power 
      laws $(k\xi)^4$ and $(k\xi)^{-2.5}$ for comparison.}
   \label{collapse}
\end{figure}
%
%
%
Figure~\ref{collapse} shows $\xi(t)^{q} E_{\rm M}(k,t)$ versus the 
scaled variable $k\xi(t)$, with $q=0.7$, plotted for different 
values of time $t$. The data is seen to collapse 
onto a single curve given $g_{\rm M}(k\xi)$: thus the magnetic field 
evolves in a self-similar manner.
 
\section{Discussion and conclusions}

The most directly comparable simulations of decaying 3D MHD turbulence
were carried out by Biskamp and M\"uller \cite{BisMul99,MulBis00}.
They found similar results, the magnetic field evolved self-similarly, 
with a power-law behaviour at high $k$.  However, their power law was 
$k^{-5/3}$, much less steep than our $k^{-2.5}$.  There were a number of 
differences between their and our simulations:  they were able to achieve 
larger Reynolds numbers, both by having larger grids, and by using 
hyper-diffusivity (a $\nabla^4$ magnetic diffusivity term).  
However, we believe that 
the real difference is due to their initial cut-off scale being 
significantly larger, at $k_c=4$.
We have performed a run with larger initial length scale, $k_{c} = 5$. In this
case the magnetic energy spectrum develop into an approximate $k^{-5/3}$ law at
late times. However, this occurs only after the peak of the spectrum has left
the simulation box. 

There have also been renormalisation group (RG) analyses 
looking for an inverse cascade in driven MHD turbulence 
\cite{Shiromizu:1998bc,Berera:2001eb}. 
In particular Berera and Hochberg \cite{Berera:2001eb} 
saw no 
evidence for an inverse cascade. However, 
it is not clear that the results are directly 
comparable, firstly because we are considering freely decaying turbulence, 
and secondly because RG analysis can only give information about late times 
when the system is in equilibrium with the driving force.
What we have referred to as an inverse cascade is, even in the driven
case \cite{Bra01}, a time-dependent phenomenon characterised by a bump
in the power spectrum travelling to smaller wave number.

In conclusion, we find good evidence from our numerical simulations that 
helical stochastic magnetic fields show an inverse cascade
(in the sense explained above), and that even 
if only small helicity fluctuations are present initially, there is still 
weak inverse cascade.

We have determined growth laws for the magnetic and kinetic energies $E_M$ 
and $E_K$. In the helical case, \(E_M \sim t^{-0.7}\) and
\(E_K \sim t^{-1.1}\), which means that that there is no 
equipartition of energy. 
This is because the extra constraint of helicity conservation forces 
the magnetic field to transfer power to larger scales rather than 
allow it to be dissipated.  The importance of helicity is borne out by the 
fact that in the non-helical case, we find \(\xi \sim t^{0.4}\) and 
\(E_M \sim E_K \sim t^{-1.1}\).

Length scales in the magnetic field increase as \(t^{0.5}\) in the helical 
case, but slightly slower in the non-helical case, \(t^{0.4}\).  
Note that these growth laws disagree with all theoretical predictions to date 
\cite{Bis93,Ole97,Field:2000hi,Son:1999my}, which give $t^{2/3}$ in the helical case 
and $t^{2/7}$ for our power spectrum in the non-helical case.

Helical magnetic
fields are found to evolve in a self-similar way, with a scaling function  
\(g_M(z) \sim z^{-p}\) at large \(k\), where
\(p=2.5\) for \(\mathrm{Re}\sim10^2\). Note that this is significantly 
different from both the Iroshnikov-Kraichnan and 
Kolmogorov spectra, $k^{-3/2}$ and $k^{-5/3}$, respectively.

A good theoretical understanding of these scaling laws is required 
before the evolution of magnetic fields in the early Universe is 
properly understood, as a small error in the exponent makes a large 
error in the prediction of the magnetic field strength when propagated 
over many orders of magnitude in time.  For example, Vachaspati's 
contribution to these proceedings \cite{Vac01} assumes a growth law of 
$t^{2/3}$ in the length scale, based on a simple argument invoking 
helicity conservation \cite{Bis93,Son:1999my,Field:2000hi}, 
to obtain seed fields of an interesting 
strength from the electroweak transition, which is quite 
different from our observed growth law of $t^{0.5}$. 

\Acknowledgements
This work was conducted on the Cray T3E and SGI Origin platforms using COSMOS 
Consortium facilities, funded by HEFCE, PPARC and SGI. We also acknowledge
computing support from the Sussex High Performance Computing Initiative. 
MH thanks NORDITA for hospitality while this work was completed.



\begin{thebibliography}{99}
\bibitem{AckJens} MH would like to thank Jens Niemeyer for clarification of 
this point.
\end{thebibliography}
\end{document}